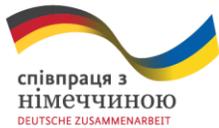 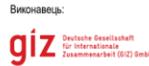 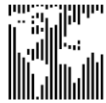

# Невідшкодування ПДВ експортерам сої або про економічні наслідки «Соєвих правок»

Олег Нів'євський, Роман Нейтер, Ольга Галиця, Павло Мартишев та Олександр Донченко, Київська школа економіки, UaFoodTrade Project

Ольга Ніколаєва, Центр аналітики зовнішньої торгівлі Trade+ при Київській школі економіки

## Історія питання

Наприкінці 2017 року[1] були прийняті зміни до Податкового кодексу України, згідно яких експортери сої та ріпаку звільнялись від сплати ПДВ. Експортери соєвих бобів (товарна позиція 1201) звільнялись від сплати ПДВ на період з 1.09.2018 року до 31.12.2021; експортери ріпаку (товарна позиція 1205) – на період з 1.01.2020 року до 31.12.2021 року.

У травні 2018 року[2] було скасовано звільнення від сплати ПДВ для підприємств – виробників соєвих бобів та ріпаку.

Задекларована мета так званих соєвих правок – збільшення переробки сої та ріпаку українськими заводами.

## Виявлена проблема політики

При експорті, для більшості товарів діє нульова ставка ПДВ. Це дозволяє уникнути подвійного оподаткування, адже у більшості країн ПДВ платиться при імпорті. При цьому експортер сплачує 0% ПДВ і отримує відшкодування ПДВ, яке він заплатив при купівлі товару.

Звільнення від сплати ПДВ для експортерів сої та ріпаку означає, що експортер не має сплачувати ПДВ, але одночасно він позбавляється можливості отримати відшкодування ПДВ, яке він сплатив за куплені ним сою та ріпак. **Таким чином звільнення від сплати ПДВ фактично стає податком на експорт сої та ріпаку**.

Більш того, очікується, що збільшення переробки сої та ріпаку українськими заводами у результаті запровадження політики має генерувати більшу додану вартість всередині країни. Однак, чи генеруватиме Україна вищу додану вартість у разі експорту продуктів переробки сої залежатиме від ефективності переробки. Неефективна переробка, навпаки, зменшує додану вартість.

Оскільки звільнення від ПДВ операцій з експорту ріпаку вступає в силу лише з наступного року, то на сьогодні можливо проаналізувати і дати економічну оцінку вже чинному звільненню від ПДВ експорту сої.

---

[1] Див. Закон України (https://zakon.rada.gov.ua/laws/show/2245-19) від 7 грудня 2017 року вніс правки до Податкового кодексу України. Серед інших змін тимчасово звільнив від сплати ПДВ

[2] Закон України 2440-VIII від 22 травня 2018 року

## Вплив політики на виробників сої

Безпосереднім наслідком дії невідшкодування ПДВ експортерам (або експортного податку) є зниження внутрішніх закупівельних цін на сою. Рисунок 1 красномовно свідчить про ціновий ефект експортного податку. Як бачимо, розрив між внутрішніми та експортними цінами суттєво збільшився починаючи з осені 2018 року – в середньому розрив збільшився з 29 до 55 дол. США/т., в той час як вартість доставки суттєво не змінилась (див. для порівняння розрив внутрішніх та експортних цін на кукурудзу на Рисунок 2). Це означає, що доходи/виручка виробників сої зменшилися в середньому на 26 дол. США/т[3]. Таким чином, враховуючи обсяг виробництва сої у 2018 році (4.5 млн т), **недоотримані доходи сільгоспвиробників сої у сукупності склали 118 млн дол. США**. Для порівняння, весь обсяг державної підтримки сільгоспвиробникам у 2019 році склав біля 6 млрд грн або 240 млн дол. США.

Звичайно, не всі виробники зазнали вищенаведених втрат доходів, а лише ті, хто не міг експортувати свій продукт самостійно. Такими підприємствами є переважно малі та середні сільськогосподарські підприємства, великі ж підприємства, як-от агрохолдинги, мають змогу безпочередньо займатися експортною діяльністю. Важко точно порахувати, який відсоток врожаю виробляється великими підприємствами, спроможними експортувати самостійно, але Рисунок 3 вказує на те, що таких підприємств переважна меншість. Ми припускаємо, що приблизно чверть врожаю сої виробляється виробниками-експортерами, віддак **недоотримані доходи сільгоспвиробників сої у сукупності склали 88,5 млн дол. США.** Більше того, великі сільгоспвиробники-експортери також отримали вигоду, оскільки могли купувати з дисконтом сою на внутрішньому ринку і реалізувати її на експорт.

У той же час, крім виробників-експортерів, нижчі закупівельні ціни є вигідними для переробників сої. У 2018/19 маркетиновому році було перероблено близько 1 млн т сої, проти 0,88 млн т попереднього маркетингового року. Це означає, що **переробники отримали додаткову вигоду (дохід) за рахунок сільгоспвиробників у розмірі 26 млн дол. США**.

---

[3] Технічний аналіз цін на сою вказує, що розрив в середньому склав 29 дол.США/т (див Додаток 2), проте в подальшому ми притримуємось більш консервативної оцінки

*Рисунок 1 Внутрішні закупівельні та експортні ціни на сою, дол. США/т*

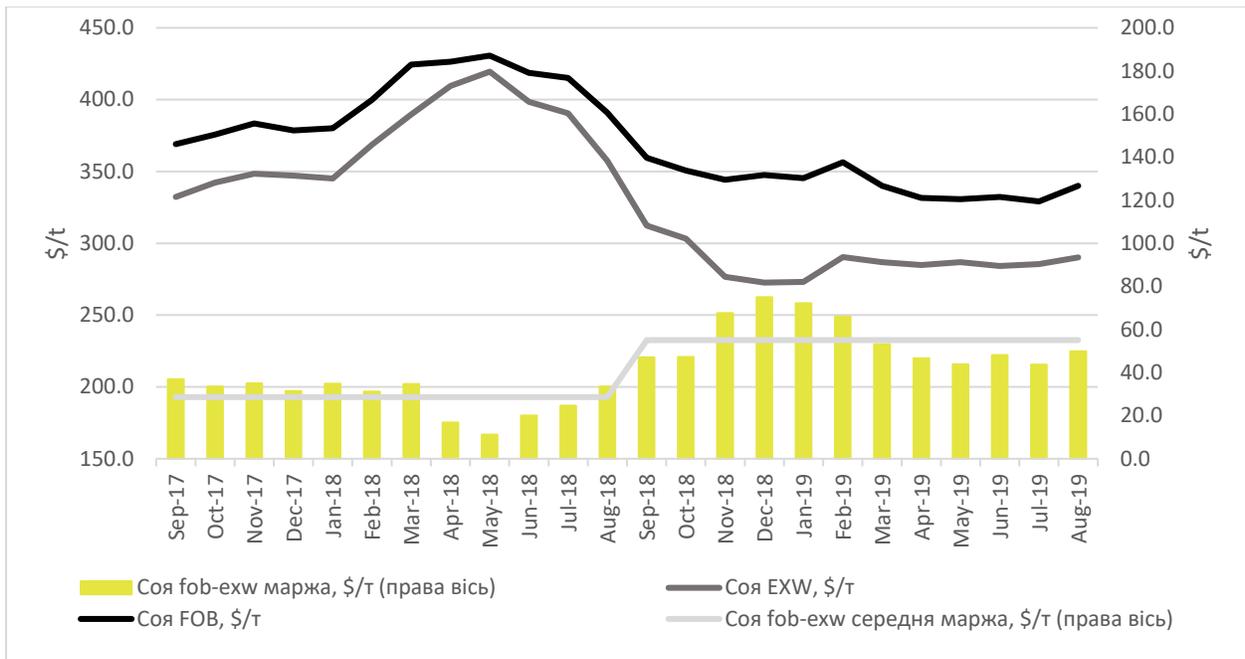

*Джерело: власна демонстрація на даних Украгроконсалт*

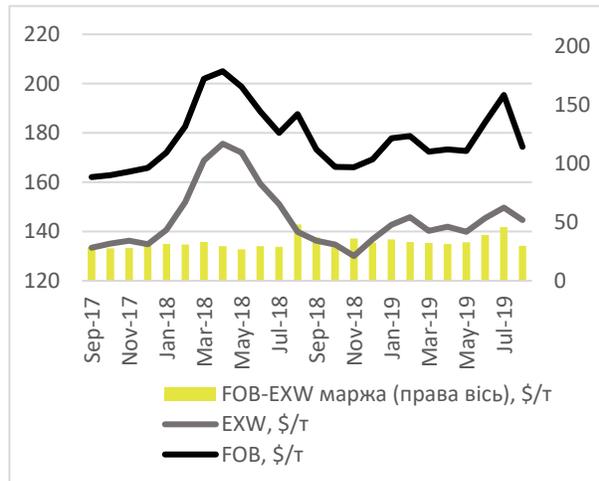

*Рисунок 2 Внутрішні закупівельні та експортні ціни на кукурудзу, дол. США/т*

*Джерело: власна демонстрація на даних Украгроконсалт*

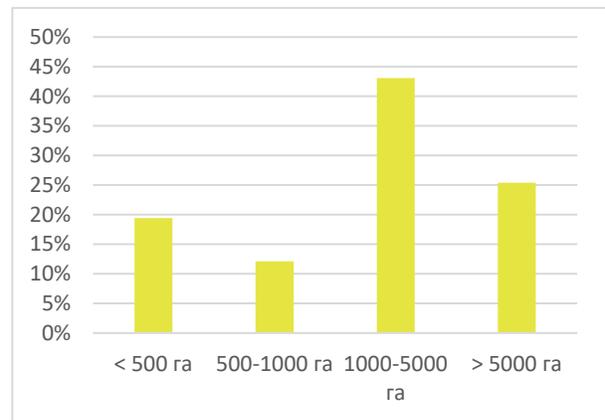

*Рисунок 3 Площі, з яких зібрано врожай сої сільгосппідприємствами різного розміру, у % до загальної площі сої за 2018 рік*

*Джерело: власна демонстрація на основі даних Укрстату*

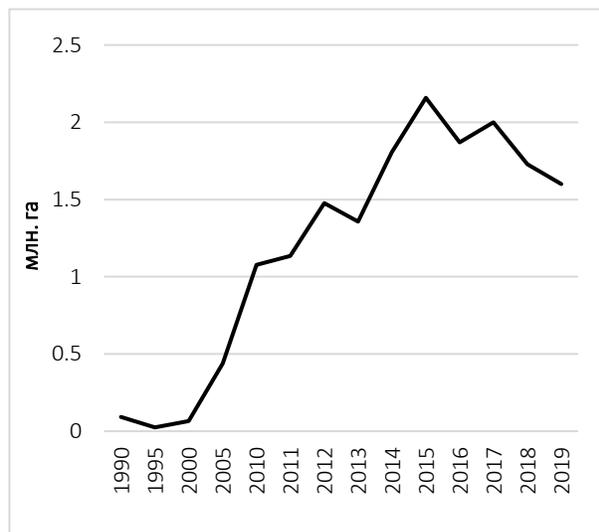

*Рисунок 4 Динаміка площ під соєю*

*Джерело: власна демонстрація на основі даних Укрстату*

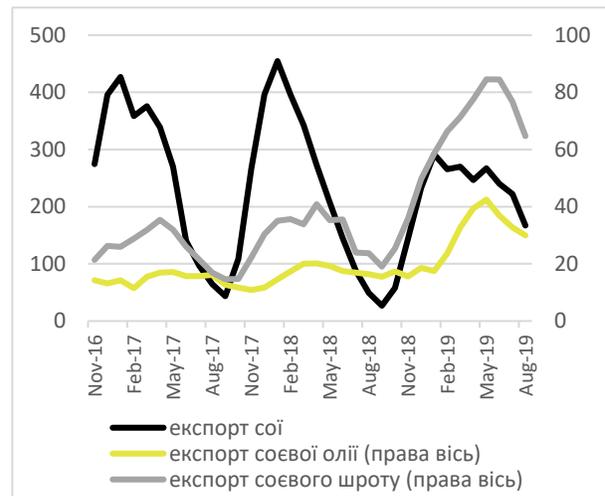

*Рисунок 5 Динаміка експорту сої та продуктів її переробки, т*

*Джерело: власна демонстрація на основі даних Укрстату*

## Вплив політики на Держбюджет в частині відшкодування ПДВ

Зміни до правил оподаткування ПДВ експортних операцій відносно сої також запроваджувались з метою зменшення зобов'язань уряду з відшкодування ПДВ експортерам; закон, який вносив зміни та доповнення до Податкового кодексу мав назву «Про внесення змін до Податкового кодексу України та деяких законодавчих актів України щодо забезпечення збалансованості бюджетних надходжень у 2018 році».

За неможливості відслідкувати зобов'язання з відшкодування ПДВ на сою за офіційними джерелами, ми розраховуємо обсяг щорічних зобов'язань з урахуванням обсягів експорту сої та продуктів її переробки, та внутрішніх цін. Як бачимо на Рисунок 5, експорт продуктів переробки суттєво зріс з часу запровадження звільнення від ПДВ експорту сої, проте експорт продуктів переробки сої не звільнявся від ПДВ, що, таким чином, автоматично збільшує зобов'язання з відшкодування ПДВ.

Беручи до уваги середні закупівельні ціни за маркетинговий рік (МР) на сою, показники експорту сої та продуктів її переробки (та вихід продуктів переробки сої), можна порахувати річні обсяги зобов'язань з відшкодування ПДВ. При цьому враховуються ПДВ зобов'язання не тільки з експорту сої, але і з експорту продуктів переробки сої, оскільки експорт продуктів переробки не звільнений від ПДВ. Як видно з Таблиця 1, обсяг оціночного номінального відшкодування ПДВ на сою та продукти його переробки в 2018/19 МР (175 млн дол. США) справді суттєво менше від 2017/18 МР (283 млн дол. США), і це за умови відшкодування ПДВ на 50%[4] всього експорту сої.

*Таблиця 1 Оціночний вплив регулювання на обсяги відшкодування ПДВ*

|  |  | 2016/2017 | 2017/2018 | 2018/2019 |
|---|---|---|---|---|
| Шрот | Ціна (EXW), $/т без ПДВ | **486** | **496** | **425** |
|  | Експорт, тис т (a) | 303 | 365 | 745 |
|  | Експорт в соєвому екв., тис т (вихід: 80%) b)=a/0.8 | 378.75 | 456.25 | 931.25 |
| Олія | Ціна (EXW), $/т без ПДВ | **787** | **795** | **703** |
|  | Експорт, тис т (c) | 177 | 192 | 324 |
|  | Експорт в соєвому екв[5]., тис т (вихід: 18%) d)=c/0.18 | 983.3 | 1066.7 | 1,800.0 |
|  | ПДВ до відшкодування, $ тис (e)=d*f*0.2 | 68,047 | 79,147 | 103,320 |
| Соя | Ціна (EXW), $/т без ПДВ (f) | **346** | **371** | **287** |
|  | Експорт, тис т (g) | 2,904 | 2,757 | 2,500 |
|  | ПДВ до відшкодування, $ тис (h)=g*f*0.2 | 200,957 | 204,569 | 143,500 |
|  | ПДВ до відшкодування (тільки на 50% експорту в 2018/19 МР), $ тис (h1)=g*f*0.2*0.5 | -- | -- | 71,750 |

---

[4] Зроблено припущення, що половина сої експортується підприємствами-виробниками, отже не відшкодовується ПДВ лише для половини екскопрту.

[5] Оскільки експорт олії в соєвому еквіваленті перебільшує експорт шроту в соєвому еквіваленті, то в подальшій оцінці сум відшкодування ПДВ при експорті сої та продуктів її переробки, ми орієнтувались лише на обсяги експорту олії в соєвому еквіваленті та безпосередньо експорту сої

| | ПДВ до відшкодування всього (соя +олія), $ тис (i)=e+h (h1 2018/19 МР) | 269,003 | 283,716 | 175,070 |
|---|---|---|---|---|
| Індекс внутр цін на сою, 1=2018/19 МР (j) | | 1.21 | 1.29 | 1.00 |
| Виробництво сої, тис т | | 3,890 | 4,461 | 3,600 |
| Індекс виробництва сої, 1=2018/19 МР (k) | | 1.08 | 1.24 | 1.00 |
| | ПДВ до відшкодування всього (індексований, в порівнюваних цінах та обсягах виробництва), $ тис (l)=i/j/k | 206,498 | 177,118 | 175,070 |

*Джерело: власні розрахунки на основі даних USDA, Украгроконсалт, Держстату*

Проте просте порівняння номінальних показників відшкодування ПДВ не враховує дві важливі речі – зміну внутрішньої ціни на сою та зміни виробництва. Зміна внутрішніх цін автоматично змінює суму ПДВ до відшкодування, навіть за незмінних обсягів експорту. Зміна виробництва є свого роду експортною базою, оскільки за збільшення виробництва весь надлишок виробництва переважно буде експортовано у вигляді сої або продуктів її переробки через доволі обмежений рівень внутрішнього споживання сої та продуктів її переробки в Україні. Тому для правильного порівня сум ПДВ до відшкодування потрібно враховувати і динаміку цін, і динаміку виробництва. А враховуючи індекс цін та виробництва, сума ПДВ до відшкодування в 2018/19 МР (175 млн дол. США) тільки фактично на 2 млн дол. США менше від показників 2017/18 МР. Якщо ж додатково врахувати ефект на внутрішню ціну від невідшкодування ПДВ (тобто врахувати індекс експортних цін на сою, а не внутрішніх), то не важко порахувати, що сума ПДВ до відшкодування в 2018/19 МР буде менше від 2017/18 МР на 18 млн дол. США.

### Сумнівний вплив на збільшення доданої вартості

Як зазначалося вище, задекларована мета соєвих правок – збільшення переробки сої та ріпаку українськими виробниками, тобто всередині країни. У результаті має збільшується додана вартість, яка генерується всередині країни. Аналізована політика нагадує практику із встановленням експортного мита на насіння соняшника. До 1999 року майже половина (в середньому до 1 млн т) насіння соняшнику експортувалась, а із запровадженням 23% експортного мита в 1999 році – експорт насіння соняшнику став економічно невигідним, тому практично все насіння йшло на внутрішню переробку[6]. Перероблі потужності олійно-екстракційних заводів за останні 16 років збільшились більш ніж у 5 разів, і зараз вони становлять близько 22 млн. т. олійних[7]. Експорт соняшникової олії продовжує зростати, і станом на сьогодні Україна посідає перше місце серед експортерів соняшникової олії у світі. Проте навіть поверхневі розрахунки ставлять під сумнів вірогідність того, що переробка соняшника дійсно генерує додаткову вартість всередині країни, – радше навпаки[8].

---

[6] Варто, правда, зазначити, що в 2001 році ставка мита була знижена до 17%, а після вступу до СОТ, ставка щорічно зменшувалась на 1 відсотковий пункт до рівня 10%, проте це вплинуло на загальну картину.

Для того щоб зрозуміти чи справді переробка сої генерує додану вартість для країни (або за яких умов генеруватиметься додана вартість), проведемо аналогічні розрахунки. На прикладі середніх експортних цін у 2018/19 маркетинговому році та виходу олії та шроту з 1 тони сої, можна побачити, за яких умов переробка генеруватиме додану вартість в країні (відповідно і фонд заробтної плати та податки), а не зменшувати її порівняно з експортом сировини, тобто чи довший ланцюжок генеруватиме більше доданої вартості від коротшого.

Розрахунки Таблиця 2 свідчать про те, що експортна виручка від реалізації 1 тони сої в середньому на 55 дол. США/т. менша від реалізації продуктів переробки цієї тони (олії та шроту). Здавалось б, все очевидно, і виручка від експорту продуктів переробки є вищою. Однак чи генеруватиме Україна вищу додану вартість у разі експорту продуктів переробки сої залежатиме від ефективності переробки. Згідно доступних даних, вартість переробки варіюється в інтервалі 30-60 дол США/т. Отже, тільки за умови собівартості переробки нижче 55 дол. США/т., переробка сої створюватиме більше доданої вартості. При собівартості переробки вище 55 дол. США/т, неефективна переробка навпаки зменшує додану вартість. Штучне зменшення закупівельної ціни на сою на 26 дол.США/т через невідшкодування ПДВ дає можливість нефективним виробникам (в кого вартість переробки більше 55 дол. США/т) продовжувати працювати, а більш ефективним виробникам отримувати додаткову вигоду. Проте важливо пам'ятати, що це відбувається за рахунок менших доходів сільгоспвиробників і це зовсім не генерує більше доданої вартості в АПК. Таким чином, зважаючи на наведені цифри, генерування додаткової доданої вартості з звільнення від ПДВ на сою – дуже спірне питання.

---

[7] https://interfax.com.ua/news/economic/533386.html
[8] http://agroportal.ua/ua/views/blogs/zapret-eksporta-lesakruglyaka-i-poshlina-na-eksport-semyan-podsolnechnika/

*Таблиця 2 Розрахунок виручки від реалізації сої та продуктів її переробки в розрахунку на 1т сої (на прикладі середніх цін 2018/19 маркетингового року)*

| Обсяг виробництва сої, т | 1 |
|---|---|
| Фактична експортна ціна сої у 2018, $/т | 332.0 |
| Виробництво соєвої олії, т (вихід олії 18%) | 0,18 |
| Фактична експортна ціна соєвої олії, $/т | 636.0 |
| Виробництво соєвого шроту, т (вихід шроту 80%) | 0,8 |
| Фактична експортна ціна соєвого шроту, $/т | 365.0 |
| Дохід від реалізації сої на експорт, $ | **332.0** |
| Дохід від реалізації соєвої олії на експорт, $ | **95.4** |
| Дохід від реалізації соєвого шроту на експорт, $ | **292.0** |
| Різниця в отриманому доході від експорту олії та шроту та експорту непереробленої сої | 55.4 |
| Приблизна вартість переробки, $/т (на основі експрес опитування переробників) | 30-60 |

*Джерело: власні розрахунки*

## Аргументи з економічної літератури

Невідшкодування ПДВ експортерам сої означає стимулювання переробки сої всередині країни, і в економічній літературі можна знайти так званий «infant industry argument» на користь такої політики. Іншими словами, держава обирає «галузь-немовля», яка на її думку має потенціал до зростання, і підтримує її певними заходами на початковому етапі, допоки вона «стане на ноги». Це виглядає добре на перший погляд, але практика запровадження такої політики у світі говорить про протилежне, а саме те, що витрати пов'язані із такою політикою перевищують вигоди[9]. Політика просування чи стимулювання розвитку нових галузей виробництва може бути виправдана у разі існування недоліків ринкових механізмів або недобросовісної конкуренції з боку закордонних урядів[10]. Жодна з цих можливих причин, однак, не виправдовує застосування такої політики стимулювання до переробки сої в Україні, яка розвивалась і до запровадження невідшкодування ПДВ (Рисунок 6 в Додатках).

---

[9] http://www.dartmouth.edu/~dirwin/docs/tinplate.pdf; https://cei.org/blog/infant-industry-argument-does-not-justify-trade-barriers
[10] https://voxukraine.org/uk/politika-prosuvannya-eksportu-zadacha-dlya-uryadu/

Проте, навіть якщо не погоджуватися з твердженням про задовільні темпи розвитку переробки сої, існує два вагомих аргументи проти подібної політики. По-перше, зазвичай держава не може ідентифікувати потенційно успішні галузі виробництва через брак глибоких знань ринків і технологічних процесів, на відміну від бізнесу. По-друге, практика визначення урядом галузей для державної підтримки несе високий ризик корупції та отримання ренти, які у свою чергу створюють значну деформацію політик, а також призводять до неефективності ринків. Можна знайти достатньо інформації про те[11], що саме це і відбулося у випадку із «соєвими правками».

Новий підхід до промислової політики[12] стимулювання розвитку та підтримки нових галузей економіки не потребує обрання державою галузей «переможців». Натомість уряди повинні працювати із бізнес-групами, промисловими лобістами, а також неурядовими організаціями шляхом їх активного залучення в обговорення й збір відгуків, створення державно-приватних проектів й проведення інтенсивної переоцінки політики. Цей процес ґрунтується на спробах і помилках, де невдача за рівнем інформативності прирівнюється до успіху, а політика підлягає постійному перегляду, з наданням особливої увагу виявленню відповідних обмежень і можливостей. Краще створити середовище, в якому компанії будь-якої сфери економічної діяльності можуть розвиватись, і ринки будуть винагороджувати успішних й карати неуспішних без участі уряду.

## Висновок

З вересня 2018 року операції з експорту сої були звільнені від сплати ПДВ («соєві правки»), що унеможливлює отримання відшкодування ПДВ при експорті сої. Проте якщо експортери є одночасно виробниками сільськогосподарської продукції, то в такому випадку вони можуть отримати відшкодування ПДВ. Попереднє експрес-дослідження дає можливість зробити наступні висновки.

1) <u>Сумнівні формальні причини запровадження політики.</u> Однією з формальних причин «соєвих правок» було стимулювання внутрішньої переробки сої за прикладом виробництва соняшникової олії (через запровадження експортного мита на соняшник). Проте:
    a. по-перше, статистичні дані свідчать про те, що переробка сої і так доволі стабільно та систематично розвивалась в Україні без

---

[11] https://www.epravda.com.ua/publications/2018/03/21/635187/
[12] https://voxukraine.org/uk/politika-prosuvannya-eksportu-zadacha-dlya-uryadu/

втручання уряду. Тому додаткове стимулювання переробки видається неаргументованим.

b. по-друге, у випадку із соняшниковим насінням (як прикладу для наслідування) навіть поверхневі розрахунки ставлять під сумнів вірогідність того, що переробка соняшника дійсно генерує додаткову вартість всередині країни.

c. по-третє, стимулювання переробки в такий спосіб є класичним прикладом підтримки «галузі-немовляти», яка, за задумом чиновників, може перерости в успішну галузь. Проте світова практика запровадження такої політики свідчить про її контрпродуктивність, оскільки витрати, пов'язані з такою політикою, зазвичай перевищують вигоду.

2) <u>Втрати сільгоспвиробників.</u> Оскільки переважна частка сої виробляється малими та середніми сільгоспвиробниками (більше 75%), які не мають змоги безпосередньо експортувати власний продукт, то таке регулювання призвело до зниження закупівельної ціни на сою в середньому на 26 дол. США/т відносно експортної ціни. За припущення, що сільгоспвиробники-експортери виробили <u>чверть врожаю</u> сої, **недоотримані доходи сільгоспвиробників сої (в основному малих та середніх) у сукупності склали 88,5 млн дол. США.**

3) <u>Вигоди переробників сої.</u> Переробники сої отримали додаткову вигоду (дохід) від заниження внутрішньої ціни на сою за рахунок сільгоспвиробників у розмірі **26 млн дол. США**.

4) <u>Бюджет/Уряд виграв від зменшення обсягу відшкодування ПДВ, але не суттєво.</u> Одним із наслідків «соєвих правок» є не тільки збільшення переробки сої в Україні, але і збільшення експорту продуктів переробки сої з України. Таким чином, соя в будь-якому випадку експортується, проте в переробленому вигляді (олія та шрот), і в цьому випадку відбувається відшкодування ПДВ. Враховуючи такий зсув експорту з сої до продуктів її переробки та враховуючи зміну цін та виробництва сої (як бази формування експорту), сума ПДВ до відшкодування в 2018/19 МР (175 млн дол. США) тільки фактично **на 2-18 млн дол. США** (в порівнювальних цінах та виробничій базі) менше від 2017/18 МР.

5) <u>Негативний сукупний ефект на економічний добробут країни.</u> Таким чином, очевидним є негативний сукупний ефект на економіку України в цілому, який оцінюється на рівні **$44,5-60,5 млн дол. США.**
6) <u>Існують ефективніші політики стимулювання.</u> Новий підхід до промислової політики стимулювання розвитку та підтримки нових галузей економіки не потребує обрання державою галузей «переможців». Натомість, ефеективнішим є створення середовища, в якому компанії будь-якої сфери промисловості можуть розвиватись, і ринки будуть винагороджувати успішних й карати неуспішних гравців без участі уряду.

# ДОДАТОК 1

*Рисунок 6 Виробництво продуктів переробки сої в Україні*

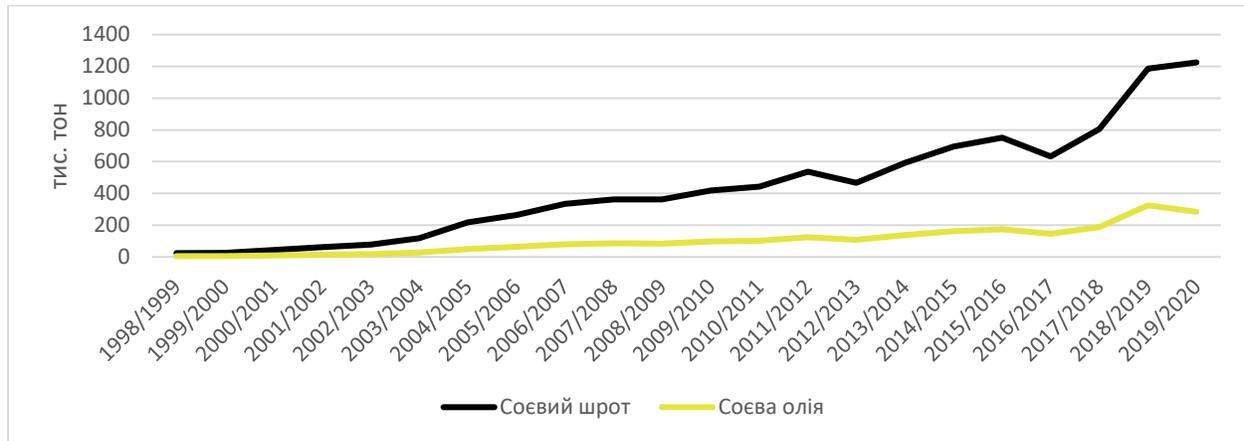

*Джерело: власна презентація на основі даних USDA. https://apps.fas.usda.gov/psdonline/app/index.html#/app/advQuery*

*Таблиця 3 Попит та пропозиція на сою та продукти її переробки, в тис. т.*

|  | 2010/11 | 2011/12 | 2012/13 | 2013/14 | 2014/15 | 2015/16 | 2016/17 | 2017/18 | 2018/19 | 2019/20 |
|---|---|---|---|---|---|---|---|---|---|---|
| **Соєвий шрот** | | | | | | | | | | |
| Початкові запаси | 20 | 32 | 0 | 28 | 94 | 115 | 46 | 27 | 46 | 64 |
| Виробництво | 442 | 537 | 466 | 593 | 695 | 751 | 632 | 806 | 1,185 | 1,225 |
| Імпорт | 40 | 25 | 7 | 3 | 2 | 2 | 2 | 3 | 3 | 2 |
| Пропозиція | 502 | 594 | 473 | 624 | 791 | 868 | 680 | 836 | 1,234 | 1,291 |
| Експорт | 5 | 5 | 15 | 80 | 216 | 347 | 303 | 365 | 745 | 725 |
| Внутр. споживання | 465 | 589 | 430 | 450 | 460 | 475 | 350 | 425 | 425 | 500 |
| Кінцеві запаси | 32 | 0 | 28 | 94 | 115 | 46 | 27 | 46 | 64 | 66 |
| **Соєва олія** | | | | | | | | | | |
| Початкові запаси | 0 | 0 | 11 | 19 | 8 | 23 | 40 | 6 | 0 | 0 |
| Виробництво | 102 | 124 | 108 | 137 | 161 | 174 | 146 | 187 | 325 | 284 |
| Імпорт | 0 | 1 | 0 | 0 | 0 | 0 | 0 | 0 | 0 | 0 |
| Пропозиція | 102 | 125 | 119 | 156 | 169 | 197 | 186 | 193 | 325 | 284 |
| Експорт | 43 | 49 | 70 | 118 | 136 | 152 | 177 | 192 | 324 | 283 |
| Внутр. споживання | 59 | 65 | 30 | 30 | 10 | 5 | 3 | 1 | 1 | 1 |
| Кінцеві запаси | 0 | 11 | 19 | 8 | 23 | 40 | 6 | 0 | 0 | 0 |
| **Соя** | | | | | | | | | | |
| Переробка | 560 | 680 | 590 | 750 | 880 | 950 | 800 | 1,020 | 1,500 | 1,550 |
| Площі | 1,037 | 1,110 | 1,411 | 1,351 | 1,800 | 2,137 | 1,858 | 1,976 | 1,729 | 1,550 |

| Початкові запаси | 150 | 103 | 0 | 99 | 265 | 166 | 113 | 151 | 21 | 113 |
|---|---|---|---|---|---|---|---|---|---|---|
| Виробництво | 1,680 | 2,264 | 2,410 | 2,774 | 3,900 | 3,932 | 4,286 | 3,890 | 4,461 | 3,600 |
| Імпорт | 2 | 1 | 2 | 3 | 4 | 5 | 7 | 8 | 7 | 6 |
| Пропозиція | 1,832 | 2,368 | 2,412 | 2,876 | 4,169 | 4,103 | 4,406 | 4,049 | 4,489 | 3,719 |
| Експорт | 989 | 1,338 | 1,323 | 1,261 | 2,422 | 2,369 | 2,904 | 2,757 | 2,500 | 1,900 |
| Внутр. споживання | 740 | 1,030 | 990 | 1,350 | 1,581 | 1,621 | 1,351 | 1,271 | 1,876 | 1,776 |
| Кінцеві запаси | 103 | 0 | 99 | 265 | 166 | 113 | 151 | 21 | 113 | 43 |
| Врожайність | 1.62 | 2.04 | 1.71 | 2.05 | 2.17 | 1.84 | 2.31 | 1.97 | 2.58 | 2.32 |

*Джерело: власна презентація на основі даних USDA*
*https://apps.fas.usda.gov/psdonline/app/index.html#/app/advQuery*

# ДОДАТОК 2. ТЕХНІЧНИЙ АНАЛІЗ ЦІН НА СОЮ

Про зростання різниці між цінами EXW за нявності і відсутності правок в середньому до 29 дол. США на тону свідчить і економетричний аналіз цін на сою. Для розрахунку була використана Баєсівська структурна модель часових рядів імплементована в мові R на основі праці Brodersen et al. (2015)[13]. Дана модель дозволяє розрахувати очікувану ціну, яка існувала б без «соєвих правок», базуючись на інших пов'язаних екзогенних змінних, проте на які прямо не вплинули законодавчі зміни. Такі екзогенні змінні – це внутрішні ціни EXW на ріпак, соняшник, FOB соя, CBOT сої зі США.

На **Ошибка! Неверная ссылка закладки.** відображено різницю між фактичною та змодельованою ціною EXW на сою в Україні в дол. США. Ціни використані в розрахунках без урахування ПДВ.

*Рисунок 7. Експортні та внутрішні (фактичні та оціночні) ціни на сою*

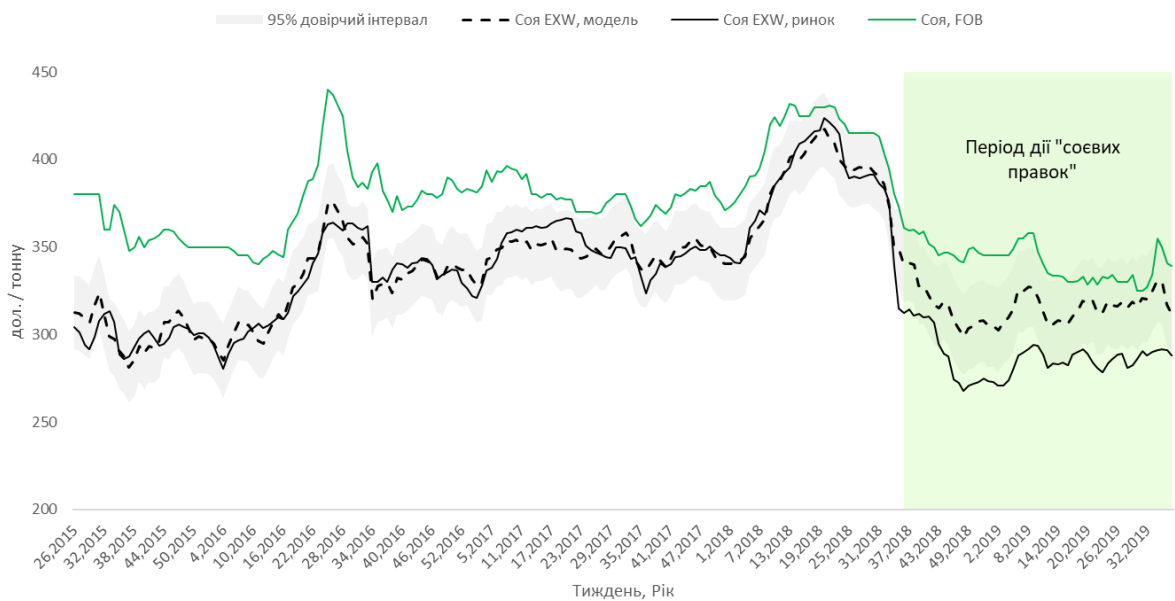

*Джерело: власні розрахунки на основі данних Ukragroconsult*

В період після «соєвих правок» з 1 вересня 2018 року середня ціна EXW в Україні на сою склала 287 дол. США. В разі відстуності штучного викривлення ринку, середня ціна EXW в Україні на сою була б на рівні 316 дол. США на тону. Довірчий інтервал на рівні 95% для модельованої ціни в разі відсутності законодавчих цін складає 304 і 327 дол. США за тонну. Різниця між фактичною EXW ціною сої на ринку і результатом моделі дає середню негативну різницю в 29 дол. США і довірчим інтервалом на рівні 95% від -17 до -40 дол. США.

---

[13] Brodersen, K. H., Gallusser, F., Koehler, J., Remy, N., & Scott, S. L. (March 01, 2015). INFERRING CAUSAL IMPACT USING BAYESIAN STRUCTURAL TIME-SERIES MODELS. The Annals of Applied Statistics, 9, 1, 247-274.